\documentclass[twocolumn,floatfix]{aastex631}

\begin{document}

\title{\textit{Astrosat} view of GX 339-4 during the peak of the recent outburst }

\author[0000-0002-9767-3927]{Shyam Prakash V. P.}
\affiliation{Space Astronomy Group, ISITE Campus, U. R. Rao Satellite Center, ISRO, Bengaluru 560037, India}

\author[0000-0002-7190-7000]{Ramadevi M.C.}
\affiliation{Space Astronomy Group, ISITE Campus, U. R. Rao Satellite Center, ISRO, Bengaluru 560037, India}

\author{Vivek K. Agrawal}
\affiliation{Space Astronomy Group, ISITE Campus, U. R. Rao Satellite Center, ISRO, Bengaluru 560037, India}







\begin{abstract}
We present the spectral and timing analyses of \textit{AstroSat} observations of the Black Hole X-ray Binary GX 339-4 when the source was close the peak of the outburst in 2024.
We find that both the spectral and timing variability of the source is indicative of it in its steep power law (SPL) state during the observations.
We used phenomenological and physical models to understand the physics and geometry of accretion during this spectral state of the source. 
Spectral fits indicate the presence of an accretion disc with a temperature of $kT\sim$0.82 keV and a hot corona with a spectral index of $\sim$2.2 along with a significant contribution from iron line emission from the accretion disc. 
Strong QPOs were detected at $\sim$4.6 Hz in the Power Density Spectra of the source along with a harmonics feature. 
Time and phase lag at the QPO frequency are studied and we find a hard lag at the QPO frequency and at the same time a soft lag at the harmonic frequency. 
We estimate the spin of the black hole and it was found that $a = 0.99 \pm 0.003$. The height of the coronal region is estimated to be about 2.5 $R_{g}$, which is found to be similar to that observed during the previous outbursts of the source.
We attempt to discuss the possible physical scenario for the observed spectral and timing features exhibited by the source.

\end{abstract}

\keywords{accretion - accretion disc - black holes - X-ray binaries - individual: GX 339-4}


\section{Introduction} \label{sec:intro}

Black hole X-ray binary systems (BHXRBs) that are transient in nature generally show outburst signatures. 
They stay in the quiescent state for a long time and exhibit outburst that can last for a few months. 
The outburst phase of a black hole X-ray binary is associated with spectral state transitions (\cite{10.1143/PTPS.155.99}, \cite{Done_2007}). 
During the outburst, the X-ray flux can go up to several orders of magnitude of that of the quiescent state (\cite{Remillard_2006}).
Typically, the source remains in the Low/Hard State (LHS) before the onset of the burst and transits to Intermediate state (IMS) during the rise of the outburst.
It is in the High/Soft (HS) state during the peak of the outburst.
This state is dominated by thermal emission processes following which the source enters the Very High State (VHS) also called the Steep Power Law (SPL) state and eventually comes back to the LHS towards the end of the outburst.
The source traces a "Q" pattern in the Hardness Intensity Diagram (HID) (\cite{Homan2001}, \cite{homan_belloni}) as the source goes through different spectral states (\cite{2004MNRAS.355.1105F}, \cite{2005A&A...440..207B}).
The relative contribution of the flux from thermal and non-thermal processes vary during spectral evolution of the source during the outburst.
(\cite{2006ARA&A..44...49R}).
  
The X-ray spectrum of BHXRBs could be described using a combination of multi-color disc blackbody and a power-law tail. 
The disc photons originating from an optically thick accretion disc are Compton up-scattered by the hot corona which leads to the formation of the hard tail. (\cite{1980A&A....86..121S}). 
Reflection features are seen in the spectra during the outburst phases of these sources (\cite{2020ApJ...899...44W}, \cite{2024AAS...24344001G}).

Understanding the power spectra and study of Quasi-Periodic Oscillations (QPO) in these binary systems is important to study the accretion flow and  geometry around compact objects. 
The origin of these features are not understood despite being known for several years. 
Black hole X-ray binary systems exhibit different types of QPOs in the PDS and these features are classified on the basis of their centroid frequency, width and amplitude (\cite{1997A&A...322..857B}, \cite{2004astro.ph.10551V}). 
Different spectral states exhibit different types of QPOs which links its association with spectral parameters. 
The quality factor is defined as (Q = $\nu/$FWHM). 
QPOs are classified into Type-A, B and C based on the quality factor, coherence and harmonic content (\cite{Wijnands_1999}, \cite{sobczak2000characterizing}). 
The most commonly seen QPOs in BHXRBs are of Type-C and are seen in LHS and VHS/SPL during the state transition. 
BHXRB systems shows Low-frequency QPOs (LFQPOs) in the range from mHz to $\sim$30 Hz (\cite{Casella}). 
The origin of these features are still under debate. Several models have been proposed in the past to explain the origin of LFQPOs such as magneto-acoustic waves (\cite{canabanc2010}) and Lense–Thirring precession model (\cite{Ingram2009}). 

GX 339-4 is classified a non-eclipsing low-mass transient X-ray binary which has shown regular outbursts in the past (\cite{Cowley_2002}).
It is known for its frequent outbursts (outburst in every 2-3 years) since its discovery in 1973 by MIT OSO-7 satellite (\cite{1973ApJ...184L..67M}) and is one of the most studied binary system in multi-wavelength. 
The latest outburst of the source started around September 2023, reached its peak intensity in February 2024. 
\cite{Hynes_2004} estimated the distance to the binary system to be between 6 and 15 kpc and the inner disc has an inclination of $i \le$ 45$^{0}$ (\cite{2002MNRAS.332..856N}). 
Many attempts were done in the past to estimate the black hole mass in the binary system and it is found to be $9^{+1.6}_{-1.2}M_{\odot}$ (\cite{Parker2016ApJ...821L...6P}) and 8.3-11.9 $M_{\odot}$ (\cite{Sreehari2019AdSpR..63.1374S}). 
The source has shown QPOs features in the PDS during the previous outbursts (\cite{2014MNRAS.438..341G}, \cite{2023ApJ...953...33J}, \cite{2023MNRAS.526.4718M}). 
The black hole has a high spin parameter ($a \ge 0.95$), estimated using spectral analysis of the broad spectrum and distorted Iron K$_{\alpha}$ line (\cite{10.1111/j.1365-2966.2008.13358.x}, \cite{10.1093/mnras/stu867}). 
The source has been extensively studied in multi-wavelength during outburst phases. 

In this paper, we discuss the results from detailed spectral and temporal studies of the source using data from \textit{AstroSat} observations on February 14 and 15, 2024, during the outburst phase of the source. 
The details of observations and data reduction steps are discussed in Section \ref{sec2_obs_DA}. 
In Section \ref{sec3_data_ana}, we describe the data analysis part and the results from spectral and temporal analysis are discussed in Section \ref{sec4_res}. 
We discuss the results from analysis and summarize the conclusions from these studies in Section \ref{sum} respectively.

\section{Observation and Data reduction}\label{sec2_obs_DA}
\subsection{AstroSat}

\begin{figure}
    \centering
    \includegraphics[width=0.47\textwidth]{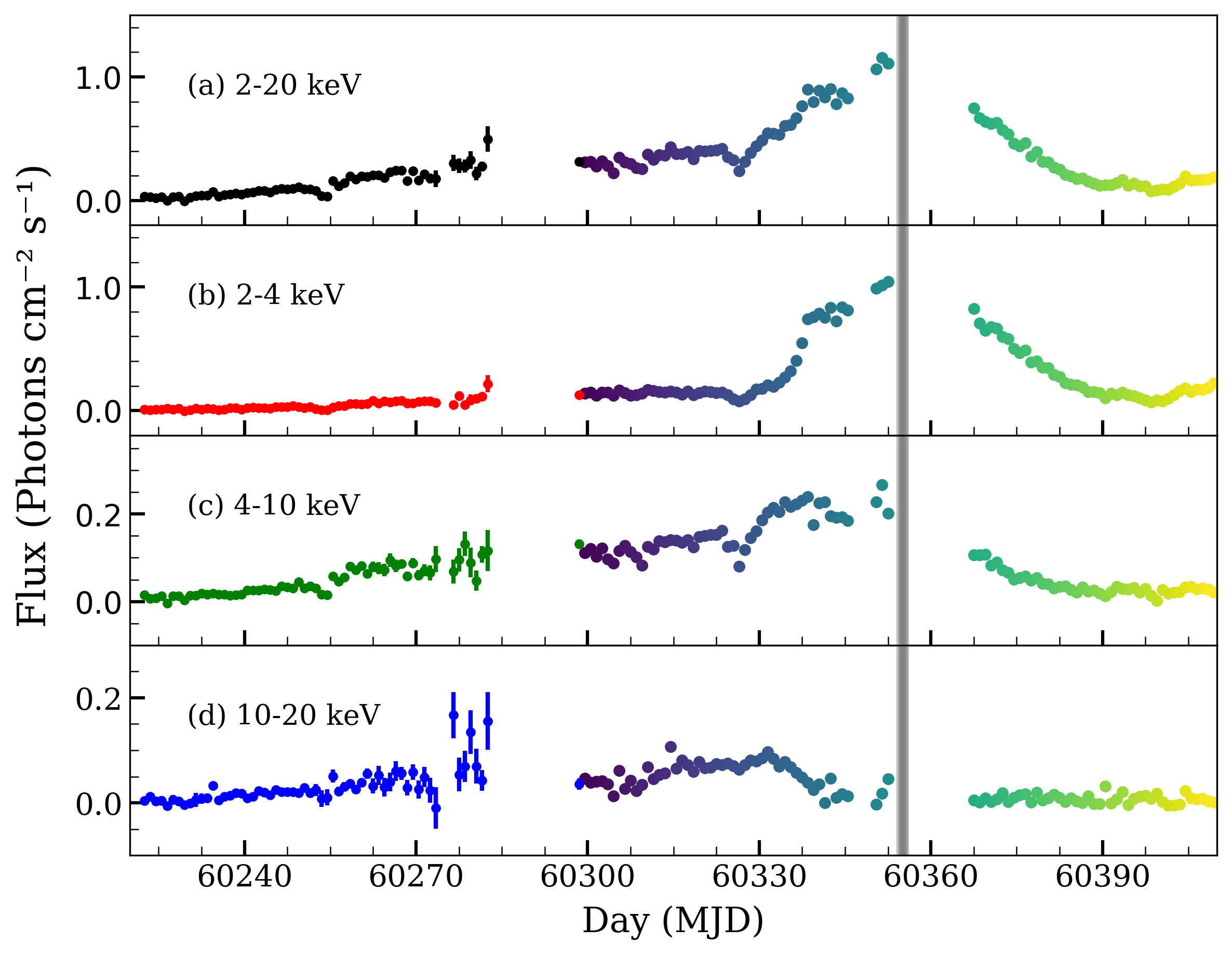}
    \caption{MAXI light curve of GX 339-4 during the 2023 outburst. The variation of (a) 2-20 keV count rate, (b) 2-4 keV count rate, (c) 4-10 keV count rate, and (d) 10-20 keV counts rates are shown. The gray shaded region marks the time period of the \textit{AstroSat} observation.}
    \label{MAXI}
\end{figure}

India's first multi-wavelength astronomy satellite \textit{AstroSat} (\cite{2014SPIE.9144E..1SS}) was launched in 2015. 
Large Area X-ray Proportional Counter (\cite{2017JApA...38...30A}) and Soft X-ray Telescope (\cite{2017JApA...38...29S})
provide data in the energy ranges 3-80 and 0.3–8 keV respectively. 
\textit{AstroSat} observed the source GX 339-4 from 2024 February 14 to 2024 February 15 during its outburst phase for an effective exposure time of $\sim$28 ks. 
The source was close to the peak of the outburst during the observations as inferred from the MAXI lightcurve of the source as shown in Figure \ref{MAXI}.
The outburst profile is plotted for different energy bands in Figure \ref{MAXI}.
The detailed \textit{AstroSat} observation log is given in Table \ref{Obs_log}.
The start time and end time for each orbits are given in MJD. 
The data from the Soft X-ray Telescope (SXT, 0.3 - 8 keV) and Large Area X-ray Proportional Counter (LAXPC, 3 - 60 keV) were used for spectro-temporal analysis. 
We reduced the LAXPC data using \href{https://www.tifr.res.in/~AstroSat_laxpc/LaxpcSoft.html} {\texttt{laxpcsoftv3.4.3\_07May2022}}\footnote{\href{https://www.tifr.res.in/~AstroSat_laxpc/software.html}{https://www.tifr.res.in/\~AstroSat\_laxpc/software.html}} developed by the Tata Institute of Fundamental Research (TIFR). Level 1 data is converted to Level 2 using the software for generating scientific products. 
The software also includes calibration files and responses. 
Using the GTI file, the light curves and spectra are generated from the Level-2 files.

The orbit wise SXT data in the photon counting mode are merged using the SXT Event Merger tool\footnote{\href{https://www.tifr.res.in/~AstroSat_sxt/dataanalysis.html}{https://www.tifr.res.in/\~AstroSat\_sxt/dataanalysis.html}}. 
The merged event file is used the to extract the image in 0.3-8 keV energy range. 
Since the source is bright in soft X-rays, pile-up effects are addressed during data reduction. 
The source counts are extracted from a circular region of radius 13 arcmins. 
In order to estimate pile-up, we extract spectrum for annular regions of different inner radius and found that exclusion of 5 arcmins region from the centre is best suited. 
For data analysis, the latest version of available RMF (sxt\_pc\_mat\_g0to12.rmf) and background (SkyBkg\_comb\_EL3p5\_Cl\_Rd16p0\_v01.pha) files are used. 
The ARF file was created using the SXTARFModule tool after correcting for the vignetting effect. The source and the background spectrum along with the respective RMF and ARF files are then grouped using \texttt{grppha} tool in Heasoft.

\begin{table}		
	\renewcommand{\arraystretch}{1.0}		
	\centering		
    \caption{Details of \textit{AstroSat} observations of the source GX 339-4 during the outburst phase in 2023. The 12 orbits considered for study and their start and end time (in MJD) are also reported.}
    \label{Obs_log}	
	\begin{tabular}{lccr}
	\hline
    Observation ID & Orbit &  Start time &   End time \\
    \hline
    & 45323 & 60354.3 & 60354.6 \\
    & 45324 & 60354.6 & 60354.7 \\
    & 45326 & 60354.7 & 60354.8 \\
    & 45327 & 60354.8 & 60354.9 \\
    & 45328 & 60354.9 & 60355.0 \\
    A05\_166T01\_9000006070 & 45331 & 60355.1 & 60355.2 \\
    & 45332 & 60355.2 & 60355.3 \\
    & 45333 & 60355.3 & 60355.3 \\
    & 45337 & 60355.3 & 60355.5 \\
    & 45338 & 60355.5 & 60355.7 \\
    & 45339 & 60355.7 & 60355.8 \\
    & 45343 & 60355.9 & 60356.0 \\
    \hline
    \end{tabular}
\end{table}

\section{Data analysis}\label{sec3_data_ana}
\subsection{Light curve and Hardness Intensity Diagram}
\begin{figure}
    \centering
    \includegraphics[width=0.47\textwidth]{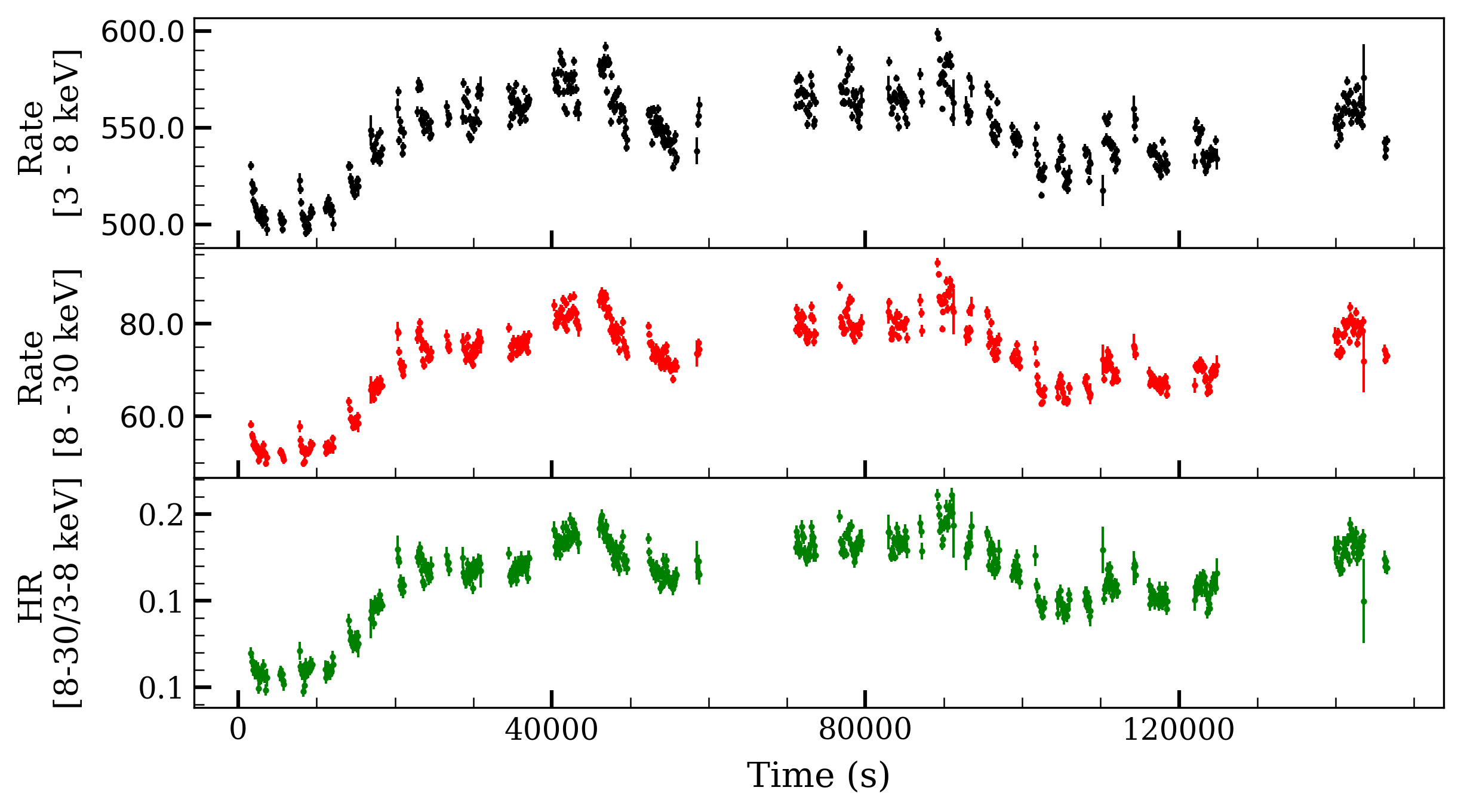}
    \caption{\textit{AstroSat}/LAXPC light curve of GX 339-4 during the 2023 outburst. The variation of (a) 8-30 keV count rate, (b) 3-8 keV count rate and (c) the hardness ratio are shown. Both the light curves shows similar behaviour.}
    \label{AstroSat/LAXPC_lc}
\end{figure}

For timing analysis, we have used data only from LAXPC20 instrument, as LAXPC10 and LAXPC30 have shown gain change effects, including gain change and gas leakage. To understand the variability in the lightcurve shown by the source during the observations, \textit{AstroSat}/LAXPC light curves in soft (3-8 keV) and hard (8-20 keV) are plotted using data from the LAXPC20 instrument and is shown in Figure \ref{AstroSat/LAXPC_lc}. Both the soft and hard bands show similar light curve profile as can be seen in Figure \ref{AstroSat/LAXPC_lc}.

\begin{figure}
    \centering
    \includegraphics[width=0.47\textwidth]{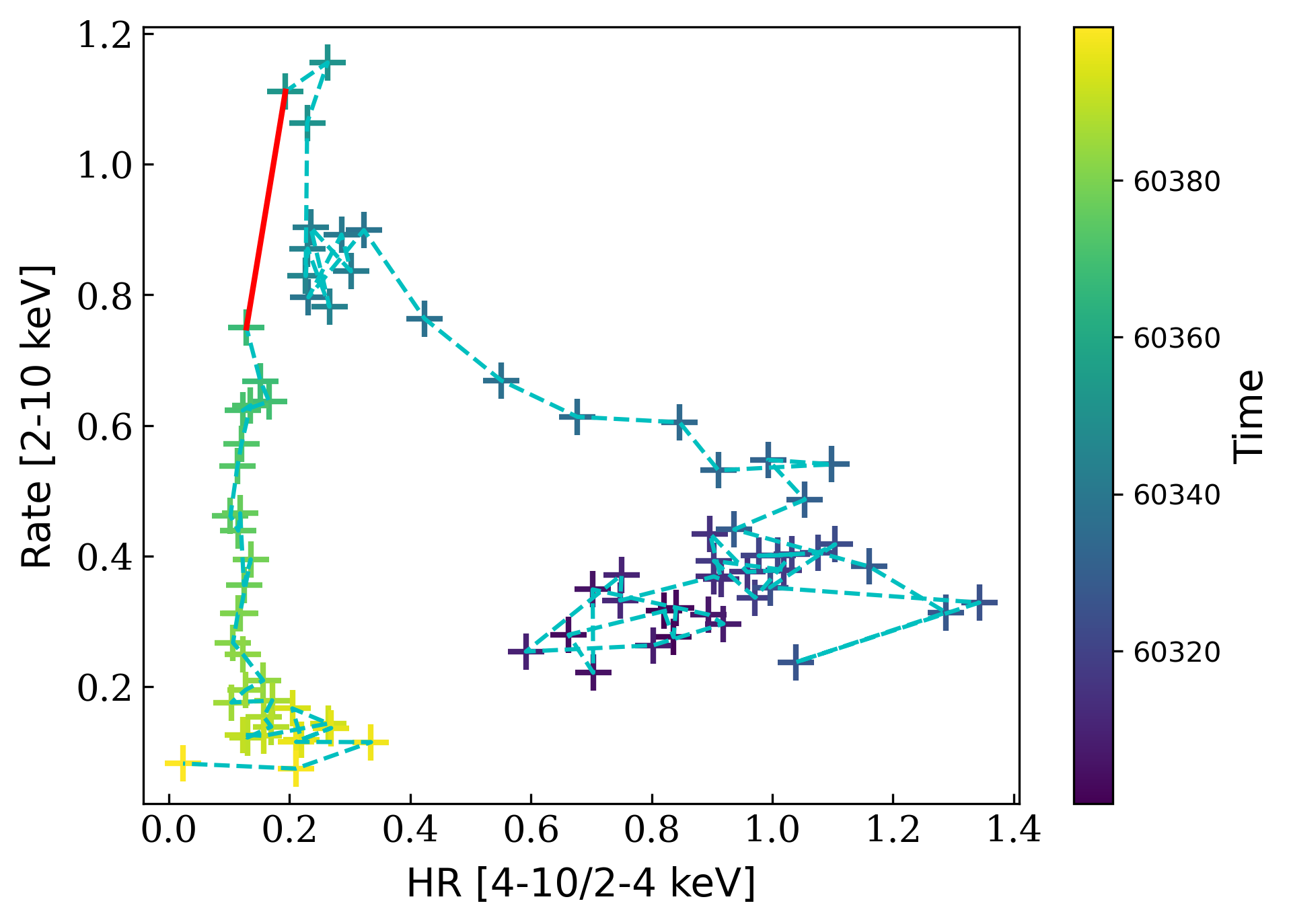}
    \caption{MAXI Hardness-intensity diagram of GX 339-4 for the complete outburst is shown here. Each point in the HID corresponds to 1-day binned MAXI observation. The intensity is taken as the count rate in the 2–20 keV energy band. The hardness ratio is defined as the ratio between the count rates in the 4–10 keV band and the 2–4 keV band. The color code given to each data point represent the time of observation. Red line denotes the time interval within which \textit{AstroSat} observations were carried out.}
    \label{MAXI_Q}
\end{figure}

\begin{figure}
    \centering
    \includegraphics[width=0.47\textwidth]{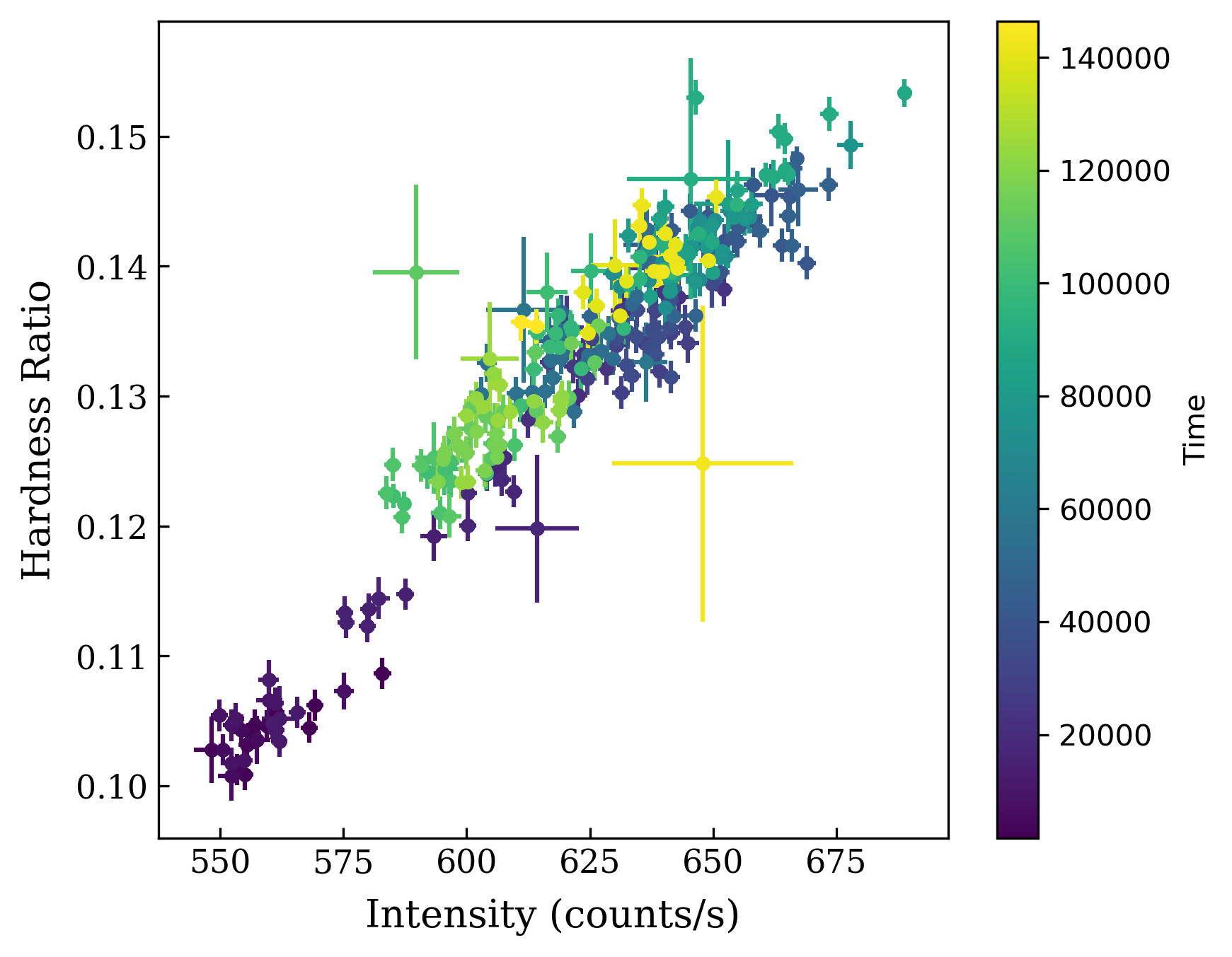}
    \caption{Hardness–intensity diagram plotted using LAXPC data during the outburst phase of the source in 2023. Each point in the HID corresponds to a binning of 100 sec. The color given to each data points are based on the observation time.}
    \label{AstroSat_HID}
\end{figure}

Hardness ratio is defined as the ratio of counts in hard band to soft band and the intensity is the sum of counts in these bands. The plot showing the variation of hardness ratio as a function of the source intensity is called the Hardness Intensity Diagram (HID). The HID plotted using the MAXI observations for the entire outburst is shown in Figure \ref{MAXI_Q}. The source traces a "Q" pattern in the HID as expected for a typical outburst of a BHXRB (\cite{Homan2001}). In Figure \ref{AstroSat_HID}, we plot the HID of the source during the outburst phase using the soft (3-8 keV) and hard (8-30 keV) bands of LAXPC data to study the changes in spectral state of the source associated with the outburst. The HID plotted for a binsize of 100 sec is shown in Figure \ref{AstroSat_HID}. The color given to each data points are based on the observation time.

\begin{figure}
    \centering
    \includegraphics[scale=0.33]{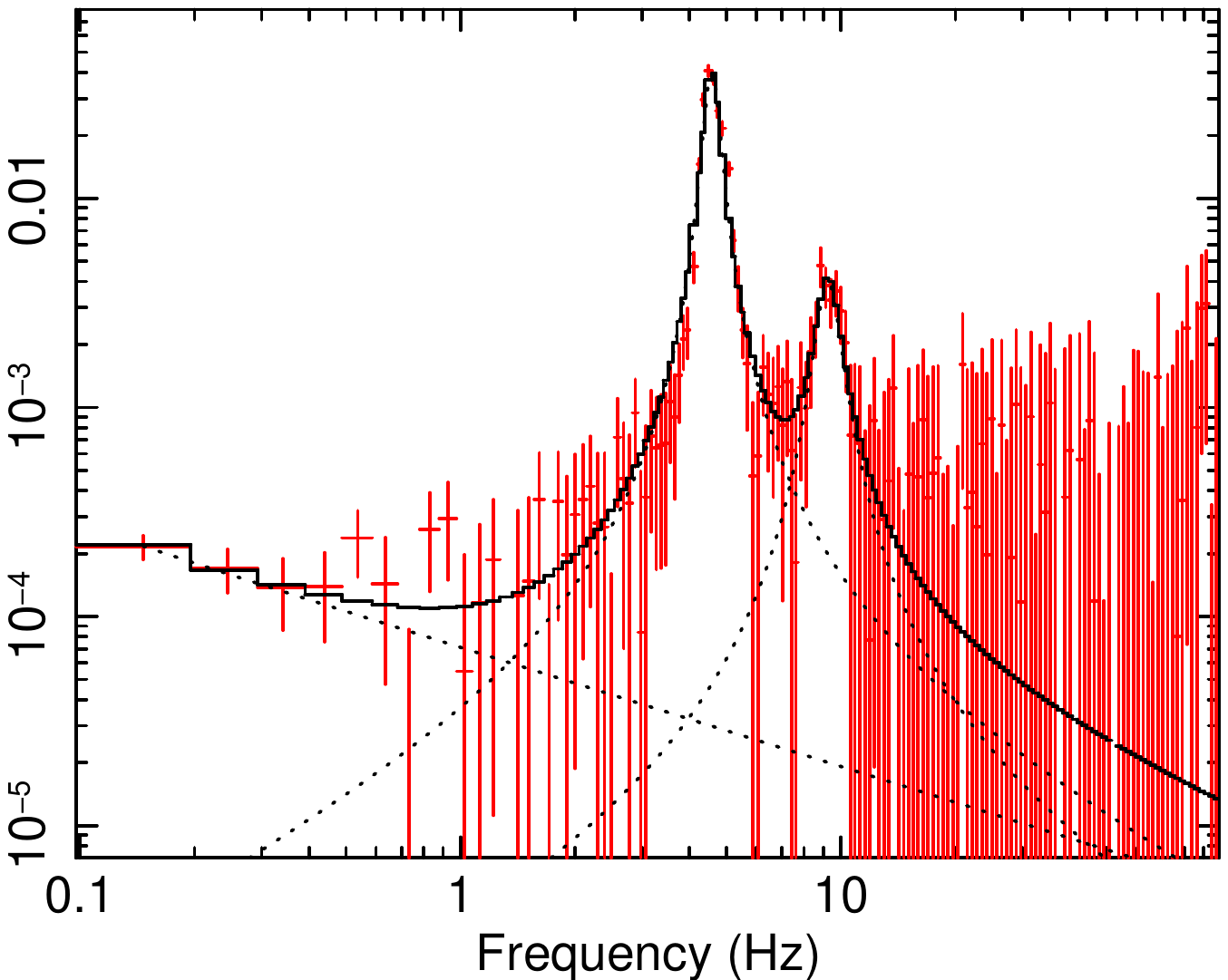}
    \caption{PDS of GX 339-4 showing QPO at $\sim$4.6 Hz and its harmonics feature at $\sim$9.4 Hz. PDS is modelled using two Lorentzian functions and a power-law function.}
    \label{pds_seg2}
\end{figure}

\subsection{Timing analysis}

QPOs are observed to be present in the Power Density Spectra (PDS) of the source during the AstroSat Observations. 
To study the evolution of QPO properties across the different orbits, we generated PDS for each orbit for the 3-50 keV energy range. 
To generate the PDS, we extracted 5 ms binned light curves in the energy range 3-50 keV using the LAXPC20 data. 
This helps in generating PDS with a maximum frequency of 100 Hz. 
We excluded frequencies above 100 Hz as they are dominated by Poisson noise.
The Leahy normalized power density spectrum is generated for data from all the orbits (\cite{1983ApJ...266..160L}). Dead-time corrected Poisson noise level is subtracted from the generated PDS (\cite{Zang1995}, \cite{deadtime_Agrawal}). 
The PDS were re-binned geometrically in the frequency space by a factor of 1.03. 
The PDS generated for each orbit is modelled using Lorentzian and power-law functions. 
The power-law model ($A\nu^{-\alpha}$) with normalization $A$ and index $\alpha$ is used to model Very Low Frequency Noise (VLFN) and Lorentzian function is defined as:

\begin{equation}
    A(E) = K(\sigma/2\pi)*[(\nu-\nu_{L})^{2} + (\sigma/2)^{2}]
\end{equation}

where $\nu_{L}$ is the centroid frequency, $\sigma$ is the FWHM of the Lorentzian peak and $K$ is the normalization factor, the square root of which gives the rms power under the Lorentzian is used to describe the QPO feature and low frequency noise in the PDS. Figure \ref{pds_seg2} shows the PDS for three orbits combined showing a QPO feature at $\sim$4.5 Hz along with a harmonics feature at $\sim$9 Hz. 

\begin{figure*}
    \centering
    \includegraphics[scale=1.5]{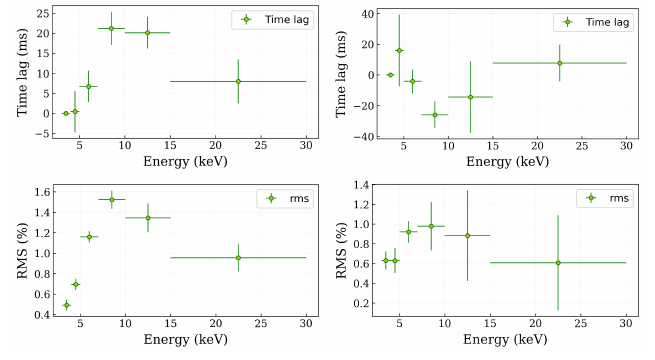}
    \caption{Energy dependent time-lag calculated for time averaged PDS during the \textit{AstroSat} observation period plotted as a function of energy in keV. Time-lag is estimated for the QPO frequency of 4.50 Hz \textit{(Left)} and harmonic feature at 9.4 Hz (\textit{Right}). The rms amplitude (in \%) calculated in the frequency range 0.1-100 Hz plotted as a function of energy.}
    \label{rms_lag-e}
\end{figure*}

We used \href{http://astrosat-ssc.iucaa.in/laxpcData}{LAXPCSoftware22Aug15} for estimating the energy dependent time-lag and rms spectra. These spectra were generated using the \texttt{laxpc\_find\_freqlag} subroutine in the software. Time-lags are calculated at the QPO frequency of 4.5 Hz and 3-4 keV as the reference energy band. The software uses cross-correlation function to obtain the phase lag and thereby the time-lag. A detailed description can be found in \cite{Nowak_1999}. The energy dependent QPO studies are carried out on the time averaged data for the complete observations. We detect no QPO features in the PDS generated for an energy range above 30 keV. The 3.0-30.0 keV energy range is divided into six bands (3-4, 4-5, 5-7, 7-10, 10-15, 15-30) to understand the dependence of QPO rms and time-lag at QPO frequency on energy. The variation in rms and time-lag as a function of energy is shown in Figure \ref{rms_lag-e}. 

\subsection{Spectral analysis}

The simultaneous SXT (0.7-7.0 keV) and LAXPC (3.0-20.0 keV) spectrum for ten orbits of the observation were used for spectral analyses.
We carried out the spectral analysis in 0.7-20 keV energy range using \texttt{XSPEC v12.10} (\cite{Arnaud1996}). 
The energy range above 20.0 keV was not considered during the spectral fitting because of the dominance of background above this energy range for LAXPC spectrum. 
To understand the emission mechanism in the source and to study the evolution of different physical parameters during the outburst, we considered spectral fitting using phenomenological and physical models. Initially, we fit the spectrum using a combination of a multi-color disk blackbody model (\textit{diskbb} \cite{1984PASJ...36..741M} \cite{1986ApJ...308..635M}) and a \textit{power law} model. 
Spectral fitting is done for each orbit in the observations. As a physical model, we use a relativistic reflection model either \textit{relxillp} or \textit{relxillcp} (\cite{2013ApJ...768..146G}) in place of \textit{power-law} along with \textit{diskbb} model to describe the spectrum in the 0.7-20.0 keV energy range. 

\subsection{Spectral models}

\begin{figure*}
    \centering
    \includegraphics[width=1\textwidth]{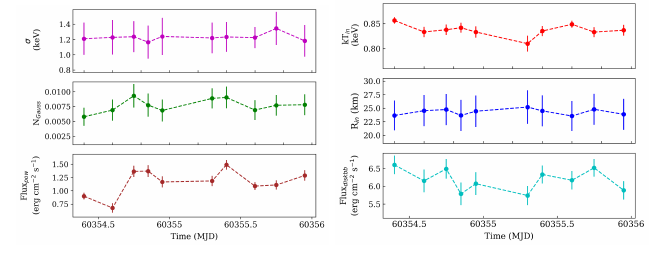}
    \caption{Evolution of key spectral parameters for the phenomenological model \textit{tbabs*(diskbb+gauss+pow)} for orbit-wise analysis. The inner disc radius is calculated from the Normalization of \textit{diskbb} model.}
    \label{Phen_model_params_binwise}
\end{figure*}

We fit the spectra using a combination of multicolor disc blackbody model (\textit{diskbb}) and a \textit{power-law} model which is referred to as Model-1. 
Strong residuals at around 6.4 keV indicated the presence of an Iron K$_{\alpha}$ line. We introduced a \textit{Gaussian} line profile to address the presence of Iron line at 6.4 keV. 
The \textit{Gaussian} line energy ($E_{L}$) is kept fixed at 6.4 keV. A systematic of 3\% was added to the data during the spectral fitting. 
The \textit{Tbabs} model in \texttt{XSPEC} was introduced to account for the galactic absorption. We set the value for Neutral Hydrogen column density, $n_{H}$ parameter free during the fitting. 
Gain correction to the SXT spectra were applied during the spectral fitting using the \texttt{gain fit} command in \texttt{XSPEC}. 
The unabsorbed flux for the disc emission and Comptonization emission were calculated in the 0.3-50.0 keV energy range using the \texttt{cflux} command in \texttt{XSPEC}. 
The total flux is calculated in the same energy range as the sum of the disc and Comptonization components.
The model describe the combined spectra from SXT and LAXPC for all the orbits with a reduced $\chi^{2}$ value close to 1 as reported in Table \ref{diskbb_fit_orb}. The evolution of the spectral parameters during the observation as a function of time (in MJD) is shown in Figure \ref{Phen_model_params_binwise}.

\begin{figure*}
    \centering
    \includegraphics[scale=1.5]{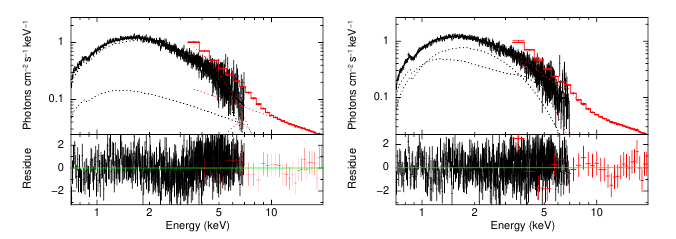}
    \caption{Combined SXT and LAXPC spectrum GX 339-4 modelled using Model-1 (\textit{tbabs*(diskbb+Gauss+pow)}) and Model-3 (\textit{tbabs*(diskbb+relxillcp)}) in the \textit{Left} and \textit{Right} panels respectively}
    \label{spec_fit_seg1}
\end{figure*}

\begin{figure}
    \centering
    \includegraphics[scale=0.6]{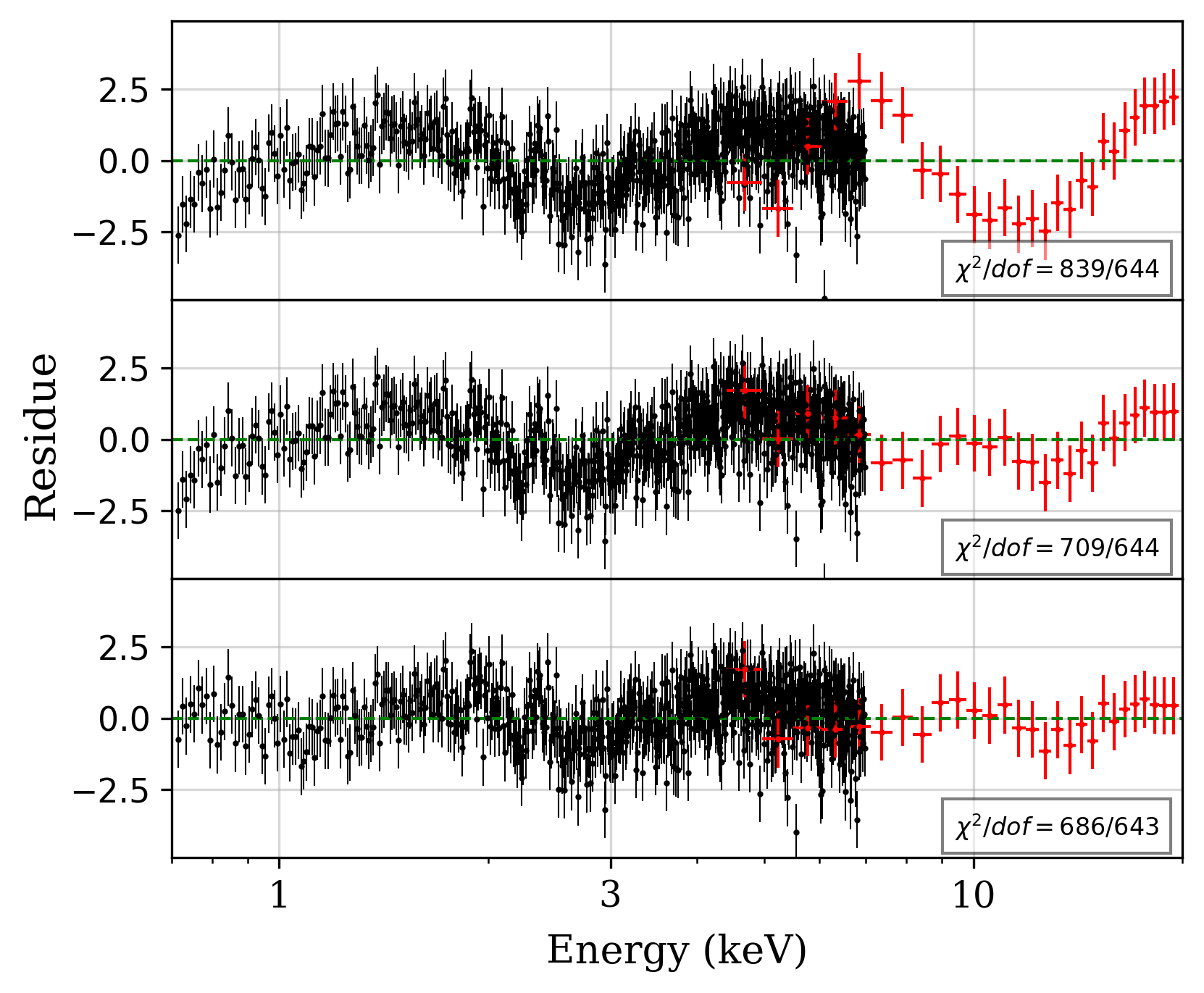}
    \caption{The residual plots for \textit{tbabs*(diskbb+pow)}, \textit{tbabs*(diskbb+pow+gauss)} and \textit{tbabs*(diskbb+relxillcp)}}
    \label{residual}
\end{figure}

In order to understand the non-thermal component in the spectra, we replaced the \textit{power-law} and \textit{Gaussian} function with \textit{relxilllp} model (\cite{García_2013}, \cite{2013MNRAS.430.1694D}) to understand the physical processes of emission in the source spectra.
The model is \textit{tbabs*(diskbb+relxilllp)} which is called the Model-2 from now on. 
This model assumes a lamp-post geometry for the corona from which low energy photons are Compton up-scattered. 
We freeze the neutral hydrogen column density ($n_{H}$) to the fit values obtained from that of Model-1. 
The spin parameter is fixed at 0.998 and the outer radius of the disc to 400 $R_{g}$ during the spectral fitting as we could not constrain these parameters and the cutoff energy was fixed at 300 keV.
A systematic of 3\% was added to the data and the Gain correction was applied as discussed above.
The spectral fit in 0.7-20 keV using Model-2 was good with $\chi^{2}$/dof $\sim$1 for all the orbits considered and the best fit parameters are reported in Table \ref{physical_model_params_orbitwise}.
We estimated the inner disc radius (in units of ISCO) and the height of the corona using spectral fit parameters from Model 2. 

We attempted to fit the spectra with a third model in which the reflection component is modelled using \textit{relxillcp} (\cite{García_2013}, \cite{2013MNRAS.430.1694D}) along with \textit{diskbb}. 
In this model (Model-3 which is \textit{tbabs*(diskbb+relxillcp)}) the inner disc radius is left free during the spectral fits. 
The best fit spectral parameters for the orbit-wise analyses are reported in Table \ref{physical_modelcp_params_orbitwise}. We could not constrain the outer disc radius and hence The outer accretion disc radius was frozen at 400 $R_g$. 
The emissivity indices for the model were fixed at the best fit values. 
The model provides a good fit for all the orbits as indicated by the $\chi^2/dof$ in Table \ref{physical_modelcp_params_orbitwise}. 
The improvements in the residuals after the introduction of the reflection component in the model is shown in the residual plot in Figure \ref{residual}.

\section{Results}\label{sec4_res}
We present the results from the detailed spectral and timing analysis of GX 339-4 using \textit{AstroSat} observations conducted when the source was close to its peak phase during the 2023 outburst. 
The results from the spectral analysis using the simultaneous SXT and LAXPC data in the 0.7–20.0 keV energy range and the evolution of timing features and their correlation with spectral parameters are discussed.

\subsection{Spectral evolution of the source}
The spectral parameters as per Model-1 indicate that source is in its VHS/SPL. There are no significant variation seen in the spectral parameters except for slight gradual changes in the disc temperature ($kT_{in}$) and flux contribution of the thermal and non-thermal components in the spectra.
Using the \textit{diskbb} normalization, we calculated the inner radius of the disc using the formula:
\begin{equation}
 Norm = \Bigg(\frac{R_{in}}{D_{10}}\Bigg)^{2} \cos\theta
\end{equation}
where the distance to the source $D_{10}$ is given in units of 10 kpc and $\theta$ is the inclination angle for the disc.
For the calculation, we assumed an inclination of 30$^{0}$ and a source distance of 8.4$\pm$0.9 kpc \cite{Parker2016ApJ...821L...6P}.
The inner disc radius is found to have an average value of $\sim$25 km during the complete period of observations.
We observe an inner disc temperature ($kT_{in}$) of $\sim$0.82 keV and a photon index ($\Gamma$) of $\sim$2.1 throughout the observation period.
No significant variation is observed in the case of both the Iron K$_{\alpha}$ line width and the inner disc radius (R$_{in}$). The R$_{in}$ is found to have an average value of 24.3 km.
The variation observed in the power-law flux follows a similar behavior to what we see in the source light curve (refer to Figure \ref{AstroSat/LAXPC_lc}). 

For spectral fitting with Model-2, the hydrogen column density ($n_{H}$) was fixed to that obtained using the Model-1 best fit parameter.
The best fit parameters for spectral fit are reported in Table \ref{physical_model_params_orbitwise}.
We see that the average value of inner disc temperature ($kT_{in}$) obtained for Model-2 ($\sim$0.81 keV) is consistent with the value obtained for Model-1.
However, the photon index ($\Gamma$) has a slightly higher value (2.5) compared to that obtained with Model 1. Also, no significant changes in any spectral parameter were found during the observations.
The \textit{relxilllp} component in Model-2 favors a softer spectrum compared to the \textit{power-law} component in Model-1.
This could be because of the fact that the \textit{relxilllp} incorporates reprocessed emission, which is absent in spectral fit with \textit{power-law} component.
The height of the corona is found to have an average value of 2.64 $R_{g}$, and the ionization parameter (log$\zeta$) and the Iron abundance ($A_{Fe}$) have values $\sim$4 and $\sim$5, respectively.

From Model-2 spectral fitting, we finds that the corona is located very close to the black hole.
Hence we attempt a spectral fit to the data using \textit{relxillcp} model replacing \textit{relxilllp} as Model-3. 
Spectral fit using Model-3 also suggest an inner disc temperature ($kT_{in}$) of $\sim$0.82 keV and a photon index of $\sim$2.2. 
Again the inner disc is found to be close to the black hole at a distance of $\sim$1.6 $R_{ISCO}$. 
The value of photon index given by the \textit{relxillcp} component in Model-3 is comparable with that using Model-1. 
The spectral fit for observations from one orbit (orbit number: 45323) using Model-1 and 3 are shown on the left, right panel respectively in Figure \ref{spec_fit_seg1}. 
All the three models provided acceptable fit for all the orbits. However, Model-3 has the best $\chi^2/dof$ value.

    \begin{table*}		
	\renewcommand{\arraystretch}{1.5}		
	\centering		
    \caption{Best-fit spectral parameter values along with their 1$\sigma$ error values for the individual orbits for 0.7-20 keV energy band combining LAXPC and SXT spectrum. The model used for fitting is \textit{tbabs*(diskbb+Gauss+pow)}.}
    \label{diskbb_fit_orb}	
	\begin{tabular}{lccccccccr} 
	\hline
	    \hline
    Orbit No. & $n_{H}$ & $kT_{in}$& $N_{diskbb}$ & $\sigma$ (keV) & $N_{Gauss} (\times 10^{-3})$ & $\Gamma$ & $N_{pow}$ & $R_{in}$ (km) & $\chi^{2}/dof$  \\
    \hline
    45323 & 0.41$^{+0.01}_{-0.01}$ & 0.86$^{+0.01}_{-0.01}$ & 686.95$^{+24.96}_{-25.71}$ & 1.21$^{+0.21}_{-0.21}$ & 5.79$^{+1.49}_{-1.55}$ & 2.16$^{+0.08}_{-0.08}$ & 0.28$^{+0.06}_{-0.07}$ & 23.60$\pm$2.75 & 839/644 \\ 

    45324 & 0.41$^{+0.01}_{-0.01}$ & 0.83$^{+0.01}_{-0.01}$ & 739.61$^{+42.63}_{-44.68}$ & 1.23$^{+0.22}_{-0.23}$ & 6.91$^{+1.75}_{-1.81}$ & 2.15$^{+0.09}_{-0.08}$ & 0.34$^{+0.07}_{-0.08}$ & 24.55$\pm$2.86 & 652/619 \\ 
    
    45326 & 0.44$^{+0.01}_{-0.01}$ & 0.83$^{+0.01}_{-0.01}$ & 752.67$^{+43.68}_{-46.19}$ & 1.23$^{+0.20}_{-0.20}$ & 9.32$^{+1.96}_{-2.03}$ & 2.14$^{+0.08}_{-0.08}$ & 0.37$^{+0.08}_{-0.09}$ & 24.76$\pm$2.87 & 608/592 \\   
    
    45327 & 0.41$^{+0.01}_{-0.02}$ & 0.84$^{+0.01}_{-0.01}$ & 687.63$^{+40.92}_{-39.39}$ & 1.16$^{+0.22}_{-0.22}$ & 7.75$^{+1.90}_{-1.99}$ & 2.21$^{+0.08}_{-0.08}$ & 0.43$^{+0.09}_{-0.11}$ & 23.67$\pm$2.86 & 681/624 \\  
    
    45328 & 0.41$^{+0.02}_{-0.02}$ & 0.83$^{+0.01}_{-0.01}$ & 733.65$^{+46.15}_{-48.73}$ & 1.24$^{+0.24}_{-0.24}$ & 6.85$^{+1.81}_{-1.88}$ & 2.17$^{+0.10}_{-0.10}$ & 0.33$^{+0.08}_{-0.10}$ & 24.44$\pm$2.90 & 678/608 \\   

    45333 & 0.43$^{+0.02}_{-0.02}$ & 0.84$^{+0.01}_{-0.01}$ & 737.98$^{+43.32}_{-41.16}$ & 1.23$^{+0.19}_{-0.20}$ & 9.04$^{+1.80}_{-1.80}$ & 2.12$^{+0.08}_{-0.08}$ & 0.32$^{+0.07}_{-0.08}$ & 24.52$\pm$2.85 & 655/623 \\
    
    45337 & 0.40$^{+0.01}_{-0.01}$ & 0.84$^{+0.01}_{-0.01}$ & 779.66$^{+69.22}_{-75.89}$ & 1.22$^{+0.20}_{-0.20}$ & 8.89$^{+1.68}_{-1.73}$ & 2.16$^{+0.08}_{-0.08}$ & 0.32$^{+0.07}_{-0.08}$ & 25.20$\pm$3.10 & 655/623 \\   
    
    45338 & 0.40$^{+0.01}_{-0.01}$ & 0.85$^{+0.01}_{-0.01}$ & 681.20$^{+27.38}_{-28.18}$ & 1.22$^{+0.08}_{-0.19}$ & 6.91$^{+1.70}_{-1.64}$ & 2.16$^{+0.08}_{-0.08}$ & 0.33$^{+0.08}_{-0.07}$ & 23.55$\pm$2.77 & 718/640 \\ 
    
    45339 & 0.40$^{+0.01}_{-0.01}$ & 0.83$^{+0.01}_{-0.01}$ & 755.17$^{+44.01}_{-41.73}$ & 1.34$^{+0.21}_{-0.22}$ & 7.72$^{+1.70}_{-1.76}$ & 2.09$^{+0.10}_{-0.09}$ & 0.26$^{+0.06}_{-0.07}$ & 24.81$\pm$2.85 & 612/613 \\   
    
    45343 & 0.40$^{+0.01}_{-0.01}$ & 0.84$^{+0.01}_{-0.01}$ & 699.28$^{+42.27}_{-44.57}$ & 1.18$^{+0.21}_{-0.21}$ & 7.81$^{+1.74}_{-1.80}$ & 2.18$^{+0.08}_{-0.07}$ & 0.37$^{+0.07}_{-0.09}$ & 23.87$\pm$2.88 & 647/613 \\  
         \hline
         \hline
    \end{tabular}
    \end{table*}

\subsection{Timing behaviour}

\begin{figure}
    \includegraphics[scale=0.45]{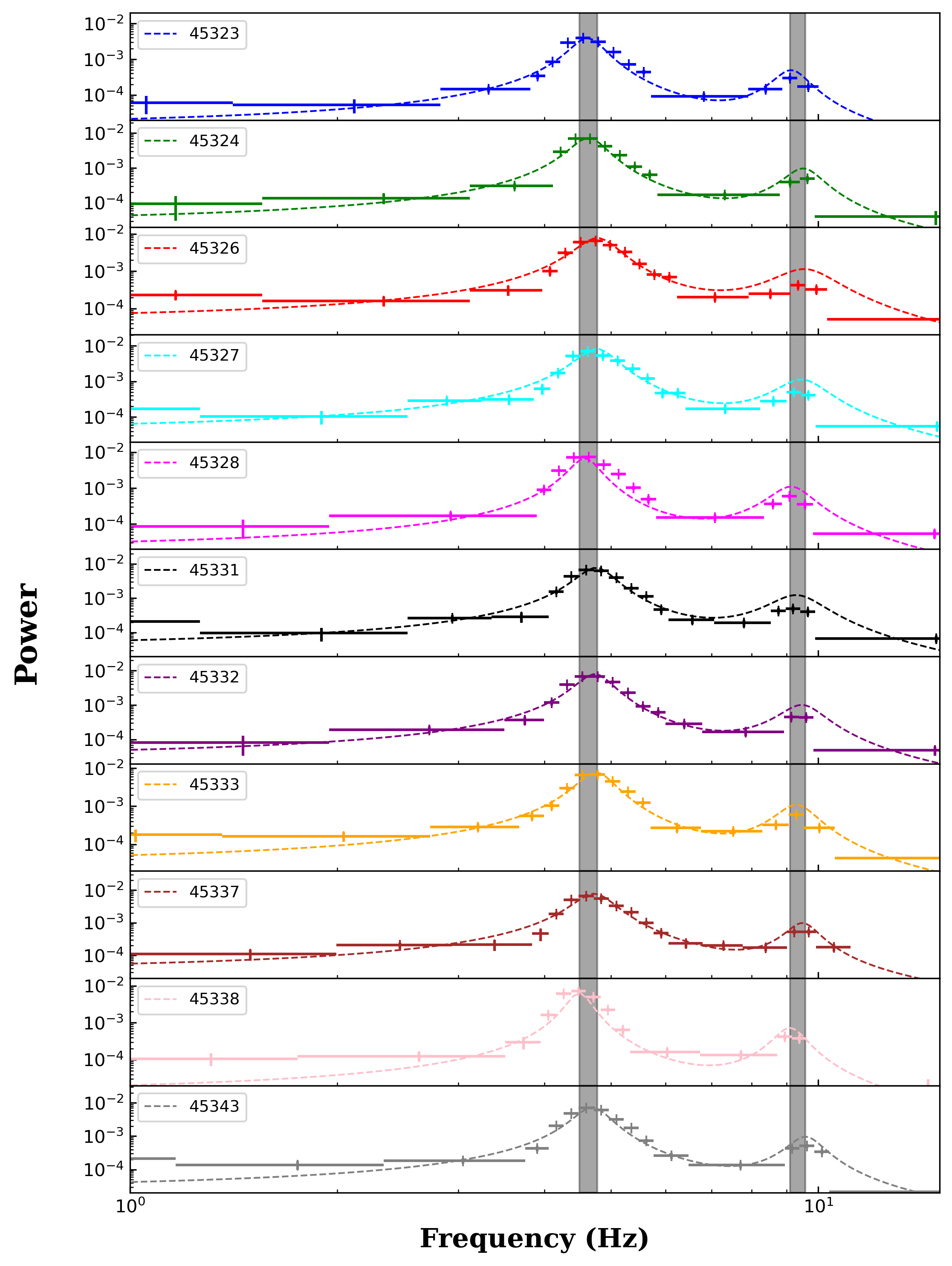}
    \caption{Orbit-wise variation of QPOs in the 3–30 keV energy band. The range of the fundamental frequency, and harmonics frequencies are marked as vertical shared regions. The 3-4 keV energy band is considered as the reference energy band.}
    \label{QPO_evol}
\end{figure}

The PDS in all orbits were fitted using a combination of Lorentzian and power-law functions.
Lorentzians were used to model the QPO, its harmonics, and the low-frequency noise in the PDS.
The best-fit parameters obtained after fitting along with the rms are reported in Table \ref{QPO_bestfit}.
We detect the presence of QPOs in all the orbits at an average frequency of 4.68 Hz, along with a harmonics feature at an average frequency of 9.36 Hz. 
The total rms variability in the PDS is found to be in the range of 0.11.
The rms strength and quality factor indicate that QPOs are C-type.
The orbit-wise timing analysis shows no significant changes in the QPO frequency throughout the observations.
Figure \ref{QPO_evol} shows the PDS generated for each orbit and the Lorentzian fit to the QPO and its harmonics.

Figure \ref{rms_lag-e} shows the results of energy-dependent timing studies conducted for the complete observations.
The rms and time-lag spectra show similar profiles.
Both plots peak around 10 keV with a hard lag profile.
The maximum of the rms and lag spectra at the QPO frequency peaks around 10 keV, suggesting that the photons with 10 keV energy are responsible for QPOs.
However, we observe a soft lag for the harmonic frequency.

    \begin{table*}		
	\renewcommand{\arraystretch}{1.5}		
	\centering		
    \caption{Best-fit spectral parameter values along with their 1$\sigma$ error values for the individual orbits for 0.7-20 keV energy band combining LAXPC and SXT spectrum. The model used for fitting is \textit{tbabs*(diskbb+relxilllp)} for individual orbits.}
    \label{physical_model_params_orbitwise}	
	\begin{tabular}{lccccccr} 
	\hline
	    \hline
         Orbit No. & $kT_{in}$ (keV) & $\Gamma$ & R$_{in}$ (ISCO) & h ($R_{g}$) & log$\zeta$ & A$_{Fe}$ & $\chi^{2}/dof$  \\
         \hline
45323 & 0.82$^{+0.01}_{-0.01}$ & 2.45$^{+0.32}_{-0.08}$ & 1.81$^{+0.24}_{-0.14}$ & 2.3$^{+0.01}_{-0.02}$ & 4.59$^{+0.02}_{-0.02}$ & 5.24$^{+0.4}_{-0.21}$ & 709/644 \\
45324 & 0.82$^{+0.01}_{-0.01}$ & 2.63$^{+0.3}_{-0.3}$ & 1.53$^{}_{-0.49}$ & 2.97$^{+0.03}_{-0.03}$ & 3.57$^{+0.1}_{-0.14}$ & 3.69$^{+1.92}_{-1.22}$ & 621/619\\
45326 & 0.82$^{+0.01}_{-0.01}$ & 2.42$^{+0.02}_{-0.02}$ & 1.53$^{+0.0}_{-0.65}$ & 2.35$^{+0.02}_{-0.02}$ & 4.3$^{+0.17}_{-0.26}$ & 5.01$^{+2.47}_{-1.23}$ & 588/624\\
45327 & 0.8$^{+0.01}_{-0.01}$ & 2.48$^{+0.04}_{-0.21}$ & 1.66$^{+0.23}_{-0.43}$ & 2.5$^{+0.01}_{-0.02}$ & 4.35$^{+0.18}_{-0.22}$ & 4.98$^{+1.93}_{-0.96}$ & 650/625\\
45328 & 0.82$^{+0.01}_{-0.01}$ & 2.6$^{+0.11}_{-0.33}$ & 1.58$^{}_{-0.58}$ & 2.98$^{+0.03}_{-0.02}$ & 3.61$^{+0.14}_{-0.14}$ & 4.99$^{+3.51}_{-1.09}$ & 666/608\\
45333 & 0.81$^{+0.01}_{-0.01}$ & 2.46$^{+0.04}_{-0.03}$ & 1.26$^{+0.16}_{-0.6}$ & 2.87$^{+0.03}_{-0.21}$ & 3.69$^{+0.17}_{-0.11}$ & 4.36$^{+1.93}_{-1.12}$ & 573/567\\
45337 & 0.79$^{+0.01}_{-0.01}$ & 2.48$^{+0.04}_{-0.05}$ & 1.68$^{+0.35}_{-0.46}$ & 2.52$^{+0.02}_{-0.01}$ & 4.16$^{+0.2}_{-0.21}$ & 4.99$^{+1.62}_{-0.53}$ & 632/624\\
45338 & 0.82$^{+0.01}_{-0.01}$ & 2.44$^{+0.02}_{-0.02}$ & 1.82$^{+0.06}_{-0.04}$ & 2.71$^{+0.02}_{-0.02}$ & 4.31$^{+0.04}_{-0.03}$ & 5.01$^{}_{-0.53}$ & 648/641\\
45339 & 0.8$^{+0.01}_{-0.01}$ & 2.49$^{+0.08}_{-0.25}$ & 1.58$^{+0.4}_{-0.46}$ & 2.57$^{+0.03}_{-0.02}$ & 3.88$^{+0.24}_{-0.29}$ & 5.0$^{+1.9}_{-0.65}$ & 630/614\\
45343 & 0.81$^{+0.01}_{-0.01}$ & 2.39$^{+0.09}_{-0.03}$ & 1.68$^{+0.55}_{-0.21}$ & 2.5$^{+0.02}_{-0.32}$ & 4.61$^{+0.2}_{-0.35}$ & 4.98$^{+1.27}_{-1.91}$ & 721/651\\
         \hline
         \hline
    \end{tabular}
    \end{table*}

        \begin{table*}		
	\renewcommand{\arraystretch}{1.5}		
	\centering		
    \caption{Best-fit spectral parameter values along with their 1$\sigma$ error values for the individual orbits for 0.7-20 keV energy band combining LAXPC and SXT spectrum. The model used for fitting is \textit{tbabs*(diskbb+relxillCp)} for individual orbits.}
    \label{physical_modelcp_params_orbitwise}	
	\begin{tabular}{lccccccr} 
	\hline
	    \hline
         Orbit No. & $kT_{in}$ (keV) & $\Gamma$ & R$_{in}$ (ISCO) & log$\zeta$ & A$_{Fe}$ & $R_{Ref}$& $\chi^{2}/dof$  \\
         \hline
45323 & 0.84$^{+0.01}_{-0.006}$ & 2.37$^{+0.02}_{-0.07}$ & 1.73$^{+0.15}_{-0.25}$ & 3.31$^{+0.07}_{-0.34}$ & 4.74$^{+0.5}_{-0.63}$ & 2.17$_{-3.36}$ & 686/643\\

45324 & 0.74$^{+0.01}_{-0.01}$ & 2.3$^{+0.01}_{-0.04}$ & 1.62$^{+0.05}_{-0.09}$ & 3.69$^{+0.13}_{-0.06}$ & 4.71$^{+0.44}_{-1.14}$ & 4.76$^{+0.0}_{-1.14}$ & 575/615\\

45326 & 0.84$^{+0.003}_{-0.002}$ & 2.06$^{+0.02}_{-0.02}$ & 1.57$^{+2.3}_{-0.65}$ & 3.36$^{+0.06}_{-0.09}$ & 4.77$_{-1.46}$ & 1.23$^{+0.08}_{-0.17}$ & 576/622\\

45327 & 0.82$^{+0.004}_{-0.002}$ & 2.21$^{+0.02}_{-0.02}$ & 1.63$^{+1.63}_{-0.38}$ & 3.38$^{+0.06}_{-0.10}$ & 4.71$^{+1.62}_{-1.06}$ & 1.78$^{+0.34}_{-0.17}$ & 636/625\\

45328 & 0.8$^{+0.006}_{-0.009}$ & 2.33$^{+0.05}_{-0.09}$ & 1.76$^{+0.11}_{-0.16}$ & 3.7$^{+0.58}_{-0.43}$ & 4.37$^{*}$ & 4.37$^{+1.65}_{-2.41}$ & 666/616\\

45333 & 0.78$^{+0.006}_{-0.006}$ & 2.3$^{+0.03}_{-0.01}$ & 1.66$^{+0.07}_{-0.14}$ & 4.06$^{+0.11}_{-0.17}$ & 4.97$^{+0.76}_{-5.09}$ & 4.34$^{+0.74}_{-0.85}$ & 559/567\\

45337 & 0.85$^{+0.002}_{-0.002}$ & 2.11$^{+0.02}_{-0.03}$ & 1.97$^{+0.43}_{-0.5}$ & 2.98$^{+0.06}_{-0.08}$ & 1.49$^{+0.84}_{-0.56}$ & 2.16$^{+0.4}_{-0.26}$ & 632/624\\

45338 & 0.83$^{+0.0}_{-0.0}$ & 2.11$^{+0.02}_{-0.02}$ & 1.87$^{+0.06}_{-0.04}$ & 3.15$^{+0.04}_{-0.03}$ & 5.0$^{+0.0}_{-0.53}$ & 3.16$^{+0.19}_{-0.44}$ & 605/641\\

45339 & 0.83$^{+0.002}_{-0.003}$ & 2.26$^{+0.02}_{-0.02}$ & 1.60$^{*}$\footnote{Frozen parameter} & 2.95$^{+0.08}_{-0.08}$ & 1.43$^{+0.59}_{-0.46}$ & 1.56$^{+0.08}_{-0.08}$ & 630/614\\

45343 & 0.81$^{+0.003}_{-0.002}$ & 2.29$^{+0.02}_{-0.02}$ & 1.65$^{+1.65}_{-0.98}$ & 4.28$^{+0.1}_{-0.12}$ & 4.77$^{*}$ & 1.89$^{+0.16}_{-0.24}$ & 623/615 \\

         \hline
         \hline
    \end{tabular}
    \end{table*}

    \begin{table*}		
	\renewcommand{\arraystretch}{1.3}		
	\centering		
    \caption{Best-fitting values for PDS fitting in 0.01-100 Hz frequency range for each orbits. The QPO and its harmonic are modelled using \textit{Lorentzian} functions.}
    \label{QPO_bestfit}	
	\begin{tabular}{lccccccccr} 
	\hline
	\hline
Orbit & $\nu_{QPO}$ (Hz) & $\sigma_{QPO}$ (Hz)& $N_{QPO}$ ($\times 10^{-3}$) & $rms_{QPO}$ (\%) & Q & $\nu_{res}$ (Hz) & $\sigma_{res}$ (Hz)& $N_{res}$ ($\times 10^{-4}$)  \\
\hline
45323 & 4.62 $\pm$ 0.02 & 0.51 $\pm$ 0.05 & 3.92 $\pm$ 0.23 & 6.26 $\pm$ 0.26 & 8.99 $\pm$ 0.88 & 9.13 $\pm$ 0.17 & 1.04 $\pm$ 0.36 & 4.84 $\pm$ 0.12  \\
45324 & 4.64 $\pm$ 0.03 & 0.54 $\pm$ 0.08 & 7.35 $\pm$ 0.64 & 8.57 $\pm$ 0.53 & 8.65 $\pm$ 1.32 & 9.51 $\pm$ 0.26 & 1.17 $\pm$ 0.85 & 9.53 $\pm$ 0.42  \\
45326 & 4.78 $\pm$ 0.03 & 0.67 $\pm$ 0.08 & 7.76 $\pm$ 0.63 & 8.81 $\pm$ 0.51 & 7.09 $\pm$ 0.88 & 9.55 $\pm$ 0.37 & 1.95 $\pm$ 0.98 & 11.06 $\pm$ 0.4 \\
45327 & 4.77 $\pm$ 0.04 & 0.62 $\pm$ 0.09 & 8.18 $\pm$ 0.69 & 9.05 $\pm$ 0.54 & 7.65 $\pm$ 1.10 & 9.46 $\pm$ 0.26 & 1.53 $\pm$ 0.64 & 10.95 $\pm$ 0.34 \\
45328 & 4.58 $\pm$ 0.02 & 0.45 $\pm$ 0.05 & 6.74 $\pm$ 0.43 & 8.21 $\pm$ 0.37 & 10.27 $\pm$ 1.05 & 9.13 $\pm$ 0.16 & 1.27 $\pm$ 0.47 & 10.82 $\pm$ 0.28 \\
45331 & 4.75 $\pm$ 0.03 & 0.59 $\pm$ 0.07 & 7.7 $\pm$ 0.57 & 8.78 $\pm$ 0.46 & 8.01 $\pm$ 0.99 & 9.29 $\pm$ 0.24 & 1.64 $\pm$ 0.59 & 11.99 $\pm$ 0.32 \\
45332 & 4.73 $\pm$ 0.03 & 0.56 $\pm$ 0.07 & 7.91 $\pm$ 0.57 & 8.9 $\pm$ 0.45 & 8.47 $\pm$ 1.00 & 9.47 $\pm$ 0.26 & 1.36 $\pm$ 0.64 & 9.84 $\pm$ 0.34  \\
45333 & 4.76 $\pm$ 0.03 & 0.59 $\pm$ 0.06 & 7.66 $\pm$ 0.5 & 8.75 $\pm$ 0.4 & 8.13 $\pm$ 0.85 & 9.29 $\pm$ 0.19 & 1.23 $\pm$ 0.45 & 10.62 $\pm$ 0.27  \\
45337 & 4.73 $\pm$ 0.04 & 0.62 $\pm$ 0.09 & 7.78 $\pm$ 0.68 & 8.82 $\pm$ 0.54 & 7.67 $\pm$ 1.13 & 9.48 $\pm$ 0.21 & 0.99 $\pm$ 0.52 & 9.52 $\pm$ 0.32  \\
45338 & 4.50 $\pm$ 0.02 & 0.37 $\pm$ 0.07 & 6.37 $\pm$ 0.64 & 7.98 $\pm$ 0.57 & 12.08 $\pm$ 2.22 & 9.10 $\pm$ 0.18 & 0.96 $\pm$ 0.43 & 7.12 $\pm$ 0.2 \\
45343 & 4.70 $\pm$ 0.03 & 0.53 $\pm$ 0.07 & 7.44 $\pm$ 0.63 & 8.63 $\pm$ 0.52 & 8.83 $\pm$ 1.23 & 9.58 $\pm$ 0.17 & 1.09 $\pm$ 0.37 & 9.42 $\pm$ 0.22  \\
         \hline
         \hline
    \end{tabular}
    \end{table*}

\section{Discussion}\label{discussion}

In this work, we report the results from a detailed spectral and timing analysis of the black hole X-ray transient GX 339-4 during its peak of the outburst phase in 2024 using \textit{AstroSat} observations.
We attempt to understand the spectral nature and the temporal variabilities shown by the source during the \textit{AstroSat} observations of the outburst.
We attempt various approaches to describe the source spectra and classify the spectral state of the source.  
The presence of type-C QPOs in the 0.1-30 Hz, the total rms power of 0.11 and a power law photon index ($\Gamma$) of $\sim$2.3 suggest that the source is in its Steep Power Law state (SPL) (\cite{Remillard_2006}). 
However, the disc to total flux ratio is slightly greater than  75\% during the observations which is about 79\% on an average. 

\subsection{Behaviour of the disc-corona system}
The spectral analysis using simultaneous LAXPC and SXT data and the QPO characteristics shows that the source was in its Steep power law state (SPL) during the \textit{AstroSat} observations.
We used different approaches to understand the spectral behavior of the source during the outburst by fitting the spectra with three different models.
In Model-1, the source spectrum is composed of photons from a multicolor disc along with that up-scattered by the hot electron corona, leading to a power-law spectrum at hard energies.
An Iron line emission is also visible.
In Model-2 and 3, we consider reflection features and reprocessed emission of disc photons by the hot corona that assumes a lamp-post geometry or a compact corona closer to the black hole respectively. Similar source behaviour has been exhibited by the source during previous outbursts (\cite{2024MNRAS.527.2128J}, \cite{2023ApJ...950....5L}).

Spectral fit using the Model-1 shows that the inner radius of the disc (calculated from the normalization of \textit{diskbb}) is very close to ISCO (1.72 $R_{ISCO}$) with a temperature of $\sim$0.81 keV and a photon index of $\sim$2.1 for the power-law component.
The observed temperature and photon index values are consistent with the spectral properties of the Steep Power Law state or the Very High State observations of the source during the previous outbursts (\cite{2009MNRAS.400.1603M}, \cite{2020ApJ...890...53S}).
The inner radius of the disc is found to be $\sim$23 km (1.72 $R_{ISCO}$).
Modeling the disc emission using the \textit{kerrbb} model helps us estimate the spin parameter for the black hole.
The spin is estimated to be $\sim$0.99 from the spectral fitting.
This estimate is consistent with the previous observations of the source, which show a highly rotating black hole in both soft and hard states (\cite{2008xru..confE..39R} and \cite{Parker2016ApJ...821L...6P} respectively).

Both Model-2 and Model-3 incorporates the reprocessed emission of the disc photons by the corona.
The Model-2 assumes the corona as a point source located at a certain height from the black hole.
The corona has a lamp-post geometry and is situated at a height in the range $\sim$2.3-3 $R_{g}$. Both the models results in a similar results as shown in Table \ref{physical_model_params_orbitwise} and \ref{physical_modelcp_params_orbitwise}.
The estimated average inner disc radius of the disc is 1.56 $R_{ISCO}$ using Model-2 and 1.71$R_{ISCO}$ using Model-3. \cite{2024MNRAS.527.2128J} has reported similar values for the outburst of the source in 2021.
The high Iron abundance value ($>4$) from both the models could be due to the metal ions that undergo radiative levitation in the inner part of the disk, as suggested by \cite{Reynolds_2012}.
The high value of the ionization parameter and Iron abundance obtained from the spectral fits here for this outburst are found to be similar to that observed during the previous outburst of the source (\cite{2024MNRAS.527.2128J}, \cite{2020ApJ...890...53S}).

\subsection{Temporal behaviour}
Timing analysis done using the LAXPC instrument on \textit{AstroSat} indicate that the variability of the light curve in the soft and hard bands shows similar behavior. 
The regions close to the compact object produce hard photons and are expected to show higher variability as compared to soft photons (\cite{1995ApJ...455..623C}). 
The hardness intensity diagram plotted using \textit{AstroSat} observations and the comparison of the period of observation in the MAXI HID shows that the source is in the Steep Power Law state. 

QPOs are observed at an average frequency of $\sim$ 4.68 Hz in all orbits of the observations. 
PDS generated in the high energy band lacks the presence of QPOs. The observed QPO frequency changes between 4.57 Hz and 4.78 Hz. 
In all the orbits, the QPO feature is also associated with a harmonic at $\sim$9.36 Hz. 
The presence of QPOs along with the spectral parameters gives us the insight that the system is in the steep power law state (\cite{INGRAM2019101524}, \cite{Motta_2016}).

We attempt an explanation for the observed QPO frequencies in the PDS in terms of the Lense-Thirring precession model (\cite{10.1111/j.1745-3933.2009.00693.x}) in which origin of the low-frequency QPO in black hole and neutron star system are attributed to the misalignment of the hot accretion flow to the spin of the compact object. In the model, the precession frequency of the inner hot flow is given by,
\begin{equation}
 v_{\text {prec }}=\frac{(5-2 \zeta) a\left(1-\left(r_i / r_o\right)^{1 / 2+\zeta}\right)}{\pi(1+2 \zeta) r_o^{5 / 2-\zeta} r_i^{1 / 2+\zeta}\left[1-\left(r_i / r_o\right)^{5 / 2-\zeta}\right]} \frac{c}{R_g} 
\end{equation}

For calculation, we consider the value of $\zeta$ to be zero as per \cite{Fragile_2007} and fixed the outer radius of the inner hot flow to be the inner disc radius ($R_{in}$) obtained from spectral fits. The spin parameter ($a$) is equated to 0.998 as obtained from the spectral fits. We estimate the radius of the hot flow to be $\sim$14 km for a precession frequency of 4.68 Hz. This also suggest the presence of a compact corona very close to the black hole.

The time lags, observed between different energy bands, provide valuable insights into the Comptonization processes occurring in these systems. 
The X-rays generated by accretion onto the compact object undergo various interactions, which result in distinct energy-dependent behaviors. 
Time lags manifest as delays in the arrival of X-ray photons at different energy bands, reflecting the physical mechanisms and geometries in the accretion flow \cite{Cui_1997}. 
The hard photons are generated after the soft photons get up-scattered by the hot corona. Here, the soft photons reach the observer first compared to the hard photons, which cause hard lags. 
Hard lags are observed at QPO frequency during the observation. The lag spectra peak at $\sim$10 keV, indicating the association of QPOs with $\sim$10 keV photons (\cite{1989ARA&A..27..517V}). 
It is reasonable to assume that the QPO is directly linked to the comptonization process in the system. 
Such behavior of the source has been observed in previous outbursts in the soft-intermediate state by \cite{2023MNRAS.526.4718M}. 
The strength of the QPO feature is also found to be peaking at $\sim$10 keV, confirming the corona and QPO association. 
Interestingly, a milli-second soft lag is observed for the harmonic feature, which again peaks at $\sim$10 keV with rms spectra similar to those of the QPO feature.

\section{Summary}\label{sum}

We present the summary of the detailed spectral and timing studies for the GX 339-4 during the peak of the outburst in 2024 using the \textit{AstroSat} observations.
The spectral and timing properties of the source during the $\sim$28 ks observation are discussed. The key findings are listed below:

\begin{enumerate}
 \item The outburst of the source started in September 2023 and lasted for almost 8 months. \textit{AstroSat} observed the source when it was close to the peak phase of the outburst February 14 and 15, 2024.
 \item During the observations, the source was seen to be in the steep power law state.
 \item The spin of the black hole is estimated, using spectral analysis, to be close to 0.99.
 \item A broad iron line of width $\sim$1.21 keV was observed in the source spectrum.
 \item We detect Type-C QPOs at frequency $\sim$4.6 Hz along with a harmonic feature at $\sim$9.4 Hz in the PDS for all the orbits of the observations. QPO and its harmonics features are modeled using Lorentzian functions.
 \item The rms and lag spectra peak around 10 keV, which associates the origin of QPO with this energy band. Only positive lags for different energies are observed with respect to the 3–4 keV energy.
 \item The spectral and timing studies indicate the presence of a compact corona close to the Black Hole. Both lamp post model and the standard relativistic reflection model explains the corona present in the Black Hole system during these observations. Better constraints of the coronal geometry can be obtained with polarisation studies carried out with IXPE observations of the source.

\end{enumerate}

\section{Acknowledgement}
We have used data from the \textit{AstroSat} mission of ISRO archived at Indian Space Science Data Centre (ISSDC). The article has used data from the SXT and the LAXPC developed at TIFR, Mumbai, and the \textit{AstroSat} POCs at TIFR are thanked for verifying and releasing the data via the ISSDC and providing the necessary software tools. Authors thank GD, SAG; DD, PDMSA and Director, URSC for encouragement and support to carry out this research. We have used data software provided by the High Energy Astrophysics Science Archive Research Centre (HEASARC), which is a service of the Astrophysics Science Division at NASA/GSFC and the High Energy Astrophysics Division of the Smithsonian Astrophysical Observatory.
\section{Data availability}
We used the \textit{AstroSat} data archived in the \href{https://astrobrowse.issdc.gov.in/astro archive/archive/Home.jsp}{Science Data Archive} of \textit{AstroSat} Mission of the Indian Space Research Organization (ISRO). 

\bibliography{GX339-4_template}{}
\bibliographystyle{aasjournal}



\end{document}